\documentclass[a4paper,10pt]{article}

\usepackage{vmargin}
\usepackage{amsfonts}
\usepackage{amsmath}
\usepackage{amssymb}  
\usepackage{graphics,graphicx,color}
\usepackage{psfrag}

\newcommand{\be}{\begin{equation}}
\newcommand{\ee}{\end{equation}}
\newcommand{\ba}{\begin{eqnarray}}
\newcommand{\ea}{\end{eqnarray}}
\newcommand{\nn}{\nonumber}
\newcommand{\R}{\mathbb{R}}
\newcommand{\C}{\mathbb{C}}
\newcommand{\N}{\mathbb{N}}
\newcommand{\CP}{\mathbb{CP}}

\newcommand{\la}{\langle}
\newcommand{\ra}{\rangle}
\newcommand{\bn}{\mathbf{n}}
\newcommand{\z}{Z}

\newcommand{\lalg}[1]{\mathfrak{#1}}  
\newcommand{\SU}{\mathrm{SU}}
\newcommand{\SO}{\mathrm{SO}}
\newcommand{\SL}{\mathrm{SL}}

\newcommand{\su}{\lalg{su}}

\newcommand{\Hom}{\mathrm{Hom}}
\newcommand{\Aut}{\mathrm{Aut}}

\DeclareMathOperator{\tr}{tr}

\setpapersize{USletter}
\setmarginsrb{16mm}{10mm}{16mm}{8mm}{12pt}{11mm}{0pt}{13mm}		 
    
\begin{document}

\sloppy
\title{\Large\bf Asymptotics of 4d spin foam models}

\author{John W. Barrett$^a$, Richard J. Dowdall$^a$, Winston J. Fairbairn$^{a,b}$, \\ Henrique Gomes$^a$, Frank Hellmann$^a$, Roberto Pereira$^{c,d}$ \footnote{e-mails: john.barrett@nottingham.ac.uk, richard.dowdall@maths.nottingham.ac.uk, winston.fairbairn@uni-hamburg.de, henrique.gomes@maths.nottingham.ac.uk, frank.hellmann@maths.nottingham.ac.uk, roberto.pereira@aei.mpg.de}\\
\\ [1mm]
\itshape{\normalsize{$^a$School of Mathematical Sciences, University of Nottingham}} \\
\itshape{\normalsize{University Park, Nottingham, NG7 2RD, UK}} \\
\\ [1mm]
\itshape{\normalsize{$^b$Department Mathematik, Universit\"at Hamburg,}} \\
\itshape{\normalsize{Bundesstra\ss e 55,  20146 Hamburg, Germany}} \\
\\ [1mm]
\itshape{\normalsize{$^c$Centre de Physique Th\'eorique}} \\
\itshape{\normalsize{Luminy, Case 907, 13288 Marseille - Cedex 9, France}} \\
\\ [1mm]
\itshape{\normalsize{$^d$MPI f\"ur Gravitationsphysik, Albert Einstein Institute,}} \\
\itshape{\normalsize{Am M\"uhlenberg 1,  14476 Potsdam, Germany}} \\}
\date{{\small\today}}
\maketitle
\vspace{-7mm}
\abstract{We study the asymptotic properties of four-simplex amplitudes for various four-dimensional spin foam models. We investigate the semi-classical limit of the Ooguri, Euclidean and Lorentzian EPRL models using coherent states for the boundary data. For some classes of geometrical boundary data, the asymptotic formulae are given, in all three cases, by simple functions of the Regge action for the four-simplex geometry.}
\vspace{2mm}

\section{Introduction}

A spin foam model is a procedure which associates an amplitude $Z(M) \in \C$ to a closed, triangulated $4$-manifold $M$.  The data for a spin model associated to $M$ is the following.
First, a map from the set of triangles of $M$ to the set of unitary, irreducible representations of a (quantum) group. Second, an assignment of state spaces to the set of tetrahedra of $M$. A state space for a tetrahedron $\tau$ is the vector space of intertwining operators between the representations assigned to the boundary of $\tau$. Finally, the last ingredient is
an assignment of a set of amplitudes to the $4$-simplexes of $M$. The amplitude
$Z(M)$ for the triangulated manifold $M$ is then given by a weighted sum over (a subset of) the set of representations and intertwining operators.

A spin foam model can be interpreted as a discretised functional integral for a large class of theories including quantum gravity. Such an interpretation relies, in particular, on the study of the semi-classical properties of the model. A key step towards the understanding of this regime is the analysis of the asymptotic behaviour of the amplitude for the $4$-simplexes when the representation labels are taken to be large.

In this paper, we summarise the results obtained in \cite{Oo,Barrett2009,Barrett2009a},  where an asymptotic analysis of the $4$-simplex amplitudes for the Ooguri model \cite{Ooguri1992c} of topological BF theory and for both Euclidean and Lorentzian versions of the EPRL model \cite{Engle2008b} of quantum gravity was performed. For an asymptotic analysis of the whole amplitude $Z(M)$ for a closed manifold $M$ of Euclidean signature see \cite{Freidel1,Freidel2}.
This paper is based on the talk given by W.J. Fairbairn at the 2nd Corf\`u summer school and workshop on quantum gravity and quantum geometry. 

\section{Four-simplex amplitudes}

A key ingredient in the formulation of a spin foam model associated to a triangulated $4$-manifold $M$ is the amplitude associated to the $4$-simplexes.  Let  $\sigma$ be a $4$-simplex of $M$. The corresponding amplitudes for the Ooguri, Euclidean and Lorentzian EPRL models are all determined by the same data associated to the boundary $\partial \sigma$ of  $\sigma$. 

\subsection{Boundary state space}

Let $\pi_k : \SU(2) \rightarrow \Aut (V_k)$, $k \in \N/2$, denote the  spin $k$ unitary, irreducible representation of the Lie group $\SU(2)$. The tetrahedra in the $4$-simplex $\sigma$ are labelled with $a = 1, 2, . . . , 5$, which implies that the couples $ab$, $a \neq b$, label the triangles of  the simplex.
Given the assignment of a spin $k_{ab}$ to each triangle of $\partial \sigma$, one can associate a state space $\mathcal{H}_{a}$ to each tetrahedron $a$ of $\partial \sigma$ given by the $\SU(2)$-invariant subspace of the tensor product of the four representations associated to the four triangles bounding the tetrahedron 
$$
 \mathcal{H}_{a} = \mathrm{Inv}_{\SU(2)} \left( \bigotimes_{b \neq a} V_{k_{ab}}\right) \cong \Hom_{\SU(2)} \left( \C \, , \, \bigotimes_{b \neq a} V_{k_{ab}} \right).
$$
The state space for the boundary of $\sigma$ is then given by
$$
\mathcal{H}_{\partial \sigma} = \bigotimes_{a} \mathcal{H}_{a},
$$ 
and the amplitude for the $4$-simplex $\sigma$ is given by a linear map $A_{\sigma} : \mathcal{H}_{\partial \sigma} \rightarrow \C$.

Posing the asymptotic problem appropriately requires to parametrise the space of four-valent intertwiners  by introducing coherent states for the spins $k$ \cite{Livine2007a}. A coherent state for the direction $\bn \in S^2 \subset \R^3$ and spin $k$ is a unit vector $\xi$ in $V_k$ defined, up to a phase, by the condition $(\mathbf{J} \cdot \bn) \, \xi = i k \, \xi$, where $\mathbf{J}$ is a three-vector whose components are the standard anti-Hermitian generators of $\su(2)$ and the dot `$\cdot$' is the 3d Euclidean inner product. In the coherent state basis, the state $\Psi_a \in \mathcal{H}_a$ associated to the tetrahedron $a$ is given by assigning coherent states to the four boundary triangles $ab$, with fixed $a$ and varying $b$, and by $\SU(2)$-averaging the four-fold tensor product using the Haar measure on $\SU(2)$
\be
\Psi_a(k , \bn) = \int_{\SU(2)} dX \, \bigotimes_{b \neq a} \, \pi_{k_{ab}}(X) \, \xi_{ab}.
\ee
The boundary state for the boundary of $\sigma$ is
$$
\Psi(k , \bn) = \bigotimes_{a} \Psi_a.
$$
The data $\{k_{ab}, \bn_{ab} \}$ specifying the boundary state up to a phase is called the boundary data.  The asymptotic formulae depend on this boundary data and certain classes of boundary data will play a paramount role in the following.

A boundary data is called {\em non-degenerate} if, for each tetrahedron $a$, the face vectors $\bn_{ab}$ corresponding to the coherent states $\xi_{ab}$ for fixed $a$ and varying $b$ span a three-dimensional space. In this case, if the four vectors $\bn_{ab}$ satisfy the closure condition 
\be
\label{closure}
\sum_{b : b \neq a} k_{ab} \bn_{ab} = 0,
\ee
they specify an embedding of the tetrahedron in three-dimensional Euclidean space, such that the vectors are the outward face normals and the $k_{ab}$ are the areas. In this way, each tetrahedron inherits a metric and an orientation but the metrics and orientations of different tetrahedra do not necessarily match. Non-degenerate boundary data for the whole $4$-simplex is said to be {\em geometric} or {\em Regge-like} if the individual tetrahedron metrics and orientations glue together consistently to form an oriented Regge-calculus positive definite 3-metric for the boundary of the 4-simplex. This is the requirement that the induced metrics on the triangles agree for both of the tetrahedra sharing any given triangle, and the induced orientations are opposite. Such boundaries satisfy the gluing constraints of \cite{Dittrich1,Dittrich2}

For geometric boundaries, one can make a canonical choice of phase for the boundary state. For this type of boundary data, there exists a unique\footnote{up to a $\mathbb{Z}_2$ lift ambiguity discussed in \cite{Barrett2009}.} set of ten $\SU(2)$ elements $g_{ab}= g_{ba}^{-1}$ which glue together the oriented geometric tetrahedra of the boundary and map the outward normal to one tetrahedron to the inward normal to the other $g_{ba} \, \bn_{ab} = - \bn_{ba}$.
From this data, one can select the phases of the coherent states by the condition
\be
\label{Regge}
\xi_{ba} =  g_{ba} J \xi_{ab},
\ee
where $J : V_k \rightarrow V_k $ is the quaternionic structure associated to the representation $k$.
The boundary state $\Psi$ with this choice of phase is called a {\em Regge state}.

\subsection{Amplitudes}

\subsubsection{Ooguri model}

The Ooguri model is a topological model corresponding to 4d BF theory with group $\SU(2)$.
The amplitude $A_{\sigma}(\Psi) \in \C$ for the $4$-simplexes of the model evaluated on a boundary state determined by the boundary data is a $15j$ symbol. Expressed in the coherent state basis it reads
\be
\label{15j}
15j(k,\bn) = (-1)^\chi \int_{\SU(2)^5} \prod_{a} \, d X_a \prod_{a < b} \la J \xi_{ab}, X^{\dagger}_a X_b \, \xi_{ba} \ra^{2k_{ab}}.
\ee
Here,  $\xi \in \C^2$ is a coherent state in the fundamental representation, $J$ is the corresponding quaternionic structure $J : \C^2 \rightarrow \C^2$; $(z_0 , z_1) \mapsto (- \bar{z_1} , \bar{z}_0)$, and the brackets $\la , \ra$ denote the Hermitian inner product on $\C^2$. The sign factor $(-1)^\chi$ is determined by the graphical calculus relating the $15j$ spin network diagram to the above evaluation. 

\subsubsection{Euclidean EPRL model}

The Euclidean EPRL model is a model of Euclidean quantum gravity with finite Immirzi parameter $\gamma$. Throughout this paper, it will be assumed that $\gamma$ is a positive real number. When  discussing the Euclidean model it will furthermore be assumed that $\gamma < 1$. 
Under this assumption, the Euclidean EPRL model is equivalent to the FK model with finite Immirzi parameter \cite{Freidel2008}.
The construction is that of a constrained topological model based on the spin cover $\SU(2) \times \SU(2)$ of the four-dimensional rotation group. The unitary, irreducible representations of the spin group are labelled by a couple of spins $(j^+,j^-) $ and act on the finite dimensional vector space $V_{(j^+,j^-)}$.
These representations factor into representations of the diagonal $\SU(2)$ subgroup as follows
\be
V_{(j^+,j^-)} \cong \bigoplus_{j = \mid j^+ - j^- \mid }^{j^+ + j^-} V_j,
\ee
with $j$ increasing in unit steps.

The Euclidean EPRL model is constructed by identifying the boundary $\SU(2)$ representation $k$ with the highest diagonal $\SU(2)$ subgroup factor of $(j^+,j^-)$, that is,  $k = j^+ + j^-$.
More precisely, the identification involves the Immirzi parameter $\gamma$ as follows
\be
\label{cE}
j^{\pm} = \frac{1}{2} (1 \pm \gamma) k.
\ee

From this identification, one can construct $\SU(2) \times \SU(2)$ intertwiners from the coherent states $\Psi_a$ associated to the boundary tetrahedra. By forming a closed diagram from these interwiners, where the contractions involve the standard symplectic inner product on the irreducible representations of $\SU(2)$, one obtains the $4$-simplex amplitude for the Euclidean EPRL model. With the convention that an element $X$ in $\SU(2) \times \SU(2)$ is written as $(X^+,X^-)$, the amplitude for the $4$-simplex $\sigma$ is given by the formula
\be
\label{E}
A^E_{\sigma}(k, \bn) = (-1)^{\chi_E} \int_{(\SU(2) \times \SU(2))^5} \, \prod_{a} d X^+_a d X^-_a \prod_{a < b} P_{ab} ,
\ee
where the propagator $P_{ab}$ yields
\be
\label{propE}
P_{ab} =  \la J \xi_{ab}, X_a^{+\dagger} X_b^+ \, \xi_{ba} \ra^{2 j^+_{ab}}  \times \la J \xi_{ab}, X_a^{-\dagger} X_b^- \, \xi_{ba} \ra^{2 j^-_{ab}},
\ee
where the spins $j^{\pm}$ are constrained by equation \eqref{cE}.
This implies that the amplitude $A^E_{\sigma}$ is an `unbalanced' square of the 15j symbol :
$$
A^E_{\sigma}(k, \bn) = 15j(\frac{1}{2} (1 + \gamma) k,\bn) \times 15j(\frac{1}{2} (1 - \gamma) k,\bn).
$$

\subsubsection{Lorentzian EPRL model}

The EPRL model is also defined for Lorentzian signature spacetimes. The model is a constrained topological model now based on the spin cover of the Lorentz group, that is, $\SL(2,\C)$ regarded as a real Lie group. The principal series of unitary, irreducible representations of $\SL(2,\C)$ are labelled by two parameters $(n,p)$, with $n$ a half-integer and $p$ a real number. These representations act in an infinite dimensional Hilbert space $V_{(n,p)}$ of homogeneous functions of two complex variables $z = (z_0,z_1) \in \C^2$. The inner product $(,)$ is defined using the standard invariant two-form $\Omega$ on $\C^2 - \{0\}$
\be
\label{inner}
\forall f,g \in V_{(n,p)}, \;\;\;\;\;\;\;\; (f,g) = \int_{\CP^1} \Omega \, \bar{f} \, g.
\ee
The integration range is the complex projective line $\CP^1$ because the combination $\Omega \bar{f} g$ has the right homogeneity to project down from $\C^2 - \{0\}$ to $\CP^1$.

These representations split into representations of the $\SU(2)$ subgroup as
\be
V_{(n,p)} = \bigoplus_{j = |n|}^{\infty} V_j,
\ee
with $j$ increasing in steps of $1$.

The Lorentzian EPRL model is constructed by assuming\footnote{This is because $V_{(n,p)}$ is isomorphic to $V_{(-n,-p)}$. Therefore it is not necessary to consider both of these representations.} that $n \ge 0$ and by identifying the boundary $\SU(2)$ representation $k$ with the lowest  $\SU(2)$ subgroup factor of $(n,p)$, that is,  $k = n$. In fact, the full prescription is the following
\be
\label{cL}
(n,p) = (k , \gamma k).
\ee

This identification leads to the embedding of the coherent states for the boundary tetrahedra into the space of $\SL(2,\C)$ intertwiners. The contraction of these intertwiners in the inner product \eqref{inner} according to the combinatorics of the appropriate spin network diagram leads to the following amplitude for the $4$-simplexes
\be
\label{L}
A^L_{\sigma}(k, \bn) = (-1)^{\chi_L} \int_{\SL(2,\C)^5} \, \prod_{a} d X_a \, \delta(X_5) \, \prod_{a < b} P_{ab}.
\ee
Here, the delta function fixes the non-compact $\SL(2,\C)$ symmetry of the amplitude and the propagator $P_{ab}$ is defined by
\vspace{-2mm}
\be
\label{propL}
P_{ab} = c_{ab} \int_{\CP^1}\, \Omega \, \la X_a^{\dagger}z,X_a^{\dagger}z\ra^{-1-ip_{ab}-n_{ab}}\la X_a^{\dagger}z,\xi_{ab}\ra^{2n_{ab}}  \la X_b^{\dagger}z,X_b^{\dagger} z\ra^{-1+ip_{ab}-n_{ab}}\la J \xi_{ba},X_b^{\dagger}z\ra^{2n_{ab}},  
\vspace{-2mm}
\ee
where $c_{ab}$ is a constant given by $c_{ab}=\frac{ (2 n_{ab} + 1) \sqrt{n_{ab}^2+p_{ab}^2}}{\pi   ( n_{ab}+ip_{ab})}$, and $(n,p)$ are constrained by equation \eqref{cL}.

\section{Asymptotic analysis}

All the above amplitudes are integral expressions in exponential form and so the asymptotic limit, where all the boundary spins are simultaneously rescaled $k_{ab} \rightarrow \lambda k_{ab}$ and taken to be large ($\lambda \rightarrow \infty$), can be analysed with stationary phase methods.

\subsection{Asymptotic problem and critical points}

\subsubsection{Ooguri model}

The scaled $15j$ symbol can be re-expressed as 
$$
15j(\lambda k,\bn) = (-1)^\chi \int_{\SU(2)^5} \prod_{a} \, d X_a \,  \exp \left( \lambda \, S_{k,\bn}[X] \right),
$$
where the action $S$ for the asymptotic problem is complex and given by
\be
\label{action}
S_{k,\bn}[X] = \sum_{a < b} 2k_{ab} \ln \, \la J \xi_{ab}, X^{\dagger}_a X_b \, \xi_{ba} \ra.
\ee
The critical points dominating the asymptotic formula are the stationary points of $S$ which are such that the real part of $S$ is maximal, that is, $\mathrm{Re} \, S = 0$.

Such critical points are determined by a closure condition \eqref{closure} and by the equation
\be
\label{veg}
X_b \bn_{ba} = - X_a \bn_{ab},
\ee
where the $\SU(2)$ action in \eqref{veg} is defined via the homomorphism to $\SO(3)$.
The closure equation is obtained by varying the action with respect to the group variables and evaluating the result on the solutions to the equation \eqref{veg}, which expresses the maximality of the real part of the action. 

\subsubsection{Euclidean EPRL model}

Since the Euclidean EPRL amplitude in the coherent state basis is a rescaled square of the $15j$ symbol, it is immediate to see that the amplitude \eqref{E} can be re-writen as
$$
A^E_{\sigma}(\lambda k,\bn) = (-1)^{\chi_E} \int_{(\SU(2) \times \SU(2))^5} \prod_{a} \, d X_a^+ \,  d X_a^- \, \exp \left( \lambda \, S_{k,\bn}[X^+,X^-]  \right),
$$
where the action $S$ is the sum of two decoupled $15j$ actions \eqref{action}
\be
S_{k,\bn}[X^+,X^-]  = S_{j_+,\bn}[X^+] + S_{j_-,\bn}[X^-], 
\ee
with the spins $j_{\pm}$ constrained as in \eqref{cE}.

Accordingly, there are now three critical point equations. A closure constraint \eqref{closure} together with the two following equations
\be
\label{Eu}
X_b^{\pm} \bn_{ba} = - X_a^{\pm} \bn_{ab}.
\ee

\subsubsection{Lorentzian EPRL model}

The Lorentzian framework is slightly different because the representation theory of the Lorentz group is more involved. Each propagator contains an internal variable, $z$, which is integrated over. Where it is necessary to distinguish these variables on the different propagators, the notation $z_{ab}$ will be used for this variable, for each $a<b$.
In the following, the combinations
 $$\z_{ab} = X_a^{\dagger}z_{ab} \quad\text{ and }\quad \z_{ba} = X_b^{\dagger}z_{ab},$$
for each $a<b$ occur frequently; this notation will be used as a shorthand.

Using this notation, the Lorentzian propagator \eqref{propL} can be written as
\be
P_{ab} = c_{ab} \int_{\CP^1} \Omega_{ab} \left( \frac{\la \z_{ba}, \z_{ba} \ra}{\la \z_{ab}, \z_{ab} \ra} \right)^{i p_{ab}}
\left( \frac{\la \z_{ab}, \xi_{ab} \ra \la J \xi_{ba}, \z_{ba} \ra}{\la \z_{ab}, \z_{ab} \ra^{1/2} \la \z_{ba}, \z_{ba} \ra^{1/2}} \right)^{2n_{ab}}, \nn
\ee
where
$$
\Omega_{ab} = \frac{\Omega}{\la \z_{ab}, \z_{ab} \ra \la \z_{ba}, \z_{ba} \ra},$$
which is a measure on $\CP^1$.
Therefore, the four-simplex amplitude can be re-expressed as follows
\be
A^L_{\sigma}(\lambda k,\bn)= (-1)^{\chi_L} \int_{(\SL(2,\C))^5}  \prod_{a} dX_a \, \delta(X_5) \int_{(\CP^1)^{10}}  \prod_{a<b} c_{ab} \, \Omega_{ab} \, \exp \left( \lambda \, S_{k,\bn}[X,z] \right). \nn
\ee
The action $S$ for the stationary problem is given by
\be
\label{staction}
S_{k,\bn}[X,z] = \sum_{a<b} n_{ab} \ln \frac{  \la \z_{ab}, \xi_{ab} \ra^2 \la J \xi_{ba}, \z_{ba} \ra ^2}{\la \z_{ab}, \z_{ab} \ra \la \z_{ba}, \z_{ba} \ra} + i  p_{ab} \ln \frac{\la \z_{ba}, \z_{ba} \ra}{\la \z_{ab}, \z_{ab} \ra} ,
\ee
where the couple $(n,p)$ is constrained according to \eqref{cL}.
Note that the first term of the action is complex and the second term is purely imaginery.

The critical points of the action are determined by a closure condition \eqref{closure}
and two spinor equations, for each $a < b$,
\be
\label{Lo}
(X_a^{\dagger})^{-1} \, \xi_{ab} = \frac{\parallel \z_{ba} \parallel}{\parallel \z_{ab} \parallel} e^{i \theta_{ab}} (X_b^{\dagger})^{-1} J \, \xi_{ba}  \;\;\; \mbox{and} \;\;\; X_a \, \xi_{ab} = \frac{\parallel \z_{ab} \parallel}{\parallel \z_{ba} \parallel} e^{i \theta_{ab}} X_b \, J \, \xi_{ba},
\ee
where $\parallel \z_{ab} \parallel$ is the norm of $\z_{ab}$ induced by the Hermitian inner product, and $\theta_{ab}$ is a phase. The closure equation is obtained by extremizing the action with respect to the group variables and evaluating the result on the solutions to the first equation in \eqref{Lo}. This equation determines the points maximizing the real part of the action. The last equation in \eqref{Lo} is obtained from the variation of the action with respect to the spinor variables $z_{ab}$.

\subsection{Geometry of the critical points}

The critical points dominating the asymptotic formula for the Ooguri and Euclidean EPRL amplitudes are determined by the same equations and therefore have the same geometric interpretation. The Lorentzian version of the EPRL model will be treated separately.

\subsubsection{Ooguri  and Euclidean EPRL models}

To understand the geometry of the critical point equations, it is illuminating to define the variables
$$
\mathbf{b}_{ab} = k_{ab} X_a \bn_{ab}.
$$
In terms of these variables, the critical point equations \eqref{closure} and \eqref{veg} become
\be
\label{trueveg}
\sum_{b : b \neq a} \mathbf{b}_{ab} = 0, \;\;\;\; \mbox{and} \;\;\;\; \mathbf{b}_{ab} = - \mathbf{b}_{ba}.
\ee
These equations define a geometric structure called a {\em vector geometry}. It is immediate to see that a vector geometry determines a $\su(2)$-valued two-form $B$ which is constant on a $4$-simplex: the variables $\mathbf{b}_{ab}$ are identified with the surface integrals of the two-form on the triangles of the $4$-simplex and the closure condition is mapped to Stokes' theorem for the constant two-form around the boundary of the tetrahedra. 


A further geometrical picture emerges if one makes restrictions on the class of boundary data. Suppose that the boundary data is such that there exists two distinct solutions to the critical point equations, that is, two solutions to  \eqref{veg} unrelated by the symmetries of the $15j$ action given by the formula $X_a'=\epsilon_aYX_a,$ with $Y\in\SU(2)$ and $\epsilon_a=\pm1$. Call these two solutions $\{ X_a^+\}$ and $\{ X_a^-\}$, with $a=1,...,5$. From this data, one can reconstruct a {\em bivector geometry} as follows.

We introduce the vector space isomorphism 
$$
\phi :  \Lambda^2(\R^4) \rightarrow \Lambda_+(\R^4) \oplus \Lambda_-(\R^4); \;\;\;\;  B \mapsto (\mathbf{b}^+, \mathbf{b}^-),
$$
decomposing any two-form over $\R^4$, or bivector,  into self-dual and anti-self-dual components. Note that $\Lambda_{\pm}(\R^4) \cong \R^3$ as vector spaces.
From the asymptotic data, one can construct the bivectors
\be
\label{bivectorsE}
B_{ab} = (\mathbf{b}^{+}_{ab}, \mathbf{b}^{-}_{ab}), \;\;\;\; \mbox{with} \;\;\;\; \mathbf{b}^{\pm}_{ab} = k_{ab} \, X_a^{\pm} \bn_{ab}.
\ee
These bivectors satisfy the following bivector geometry conditions \cite{Barrett1,Barrett2}. 

First, they are {\em simple} because $| \mathbf{b}^{+}_{ab} | =  |\mathbf{b}^{-}_{ab}|$. Second, they are {\em cross-simple} because the bivectors \eqref{bivectorsE}, with fixed $a$ and varying $b$, live in the same 3d hyperplane $N_a^{\bot}$ defined by the unit vector $N_a$. This vector is the image of the reference vector $\mathcal{N} = (1,0,0,0)$ of the three-sphere $S^3$ under the action of the $\SU(2) \times \SU(2)$ elements $(X_a^+,X_a^-)$. Hence, the following equation holds
$$
N_{aI} B^{IJ}_{ab} = 0, \;\;\;\; \mbox{with} \;\;\;\; \gamma_E(N_a) = X_a^+ X_a^{- \dagger},
$$
where $\gamma_E: S^3 \rightarrow \SU(2)$ is the standard diffeomorphism identifying the space of unit vector $S^3 \subset \R^4$ with the unitary group $\SU(2)$.
Furthermore, the constructed bivectors satisfy {\em closure} and {\em orientation}
$$
\sum_{b:b \neq a} B_{ab} = 0 \;\;\;\; \mbox{and} \;\;\;\; B_{ab} = - B_{ba},
$$
because of the closure condition \eqref{closure} and the equations \eqref{veg} satisfied by the critical points. Under the assumption of non-degeneracy of the boundary data, these bivectors also satisfy a {\em tetrahedron} condition. Finally, one can show that the critical points determine normals $N_a$ which are either such that at least three out of the five normals are linearly independent, or such that all are pointing in the same direction. In the first case, which occurs when the two solutions $\{ X_a^+\}$ and $\{ X_a^-\}$ are distinct, the corresponding bivectors satisfy the {\em non-degeneracy} condition of a bivector geometry.

Therefore, the bivector geometry theorem \cite{Barrett1,Barrett2} implies that, if the boundary data is such that there exists two distinct solutions to the critical point equations, the bivectors \eqref{bivectorsE} are equal, up to a sign, to the bivectors of a geometric $4$-simplex in $\R^{4}$. This geometric $4$-simplex is determined up to inversion through the origin. 
Hence, a distinct pair of solutions to the critical point equations \eqref{veg} is equivalent to a geometric $4$-simplex in $\R^4$, up to inversion.

\subsubsection{Lorentzian EPRL model}

Here, we make the assumption that the boundary data is such that the critical point equations  \eqref{Lo} admit a non-trivial solution $\{X_a\}$. The geometry of the critical points is then based on the identification between spinors and null vectors. Let $\gamma_L : \R^{3,1} \rightarrow \mathbb{H}$ be the isomorphism between Minkowski space $\R^{3,1}$ and the space of $2 \times 2$ hermitian matrices $\mathbb{H}$. Call $\mathbb{H}_0^+$ the subset defined by 
$$
\mathbb{H}_0^+ = \lbrace h \in \mathbb{H} \mid \det h = 0, \;\; \mbox{and} \;\; \mathrm{Tr} \; h >0 \rbrace.
$$
The isomorphism $\gamma_L$ identifies the future null cone $C^+$ in Minkowski space with $\mathbb{H}_0^+$ because $\det \gamma_L(x) = - \eta(x,x)$, where $\eta$ is a Minkowski metric with signature $-+++$.
Therefore, using the standard map between spinors and elements of $\mathbb{H}_0^+$
$$
\zeta : \C^2 \rightarrow \mathbb{H}_0^+, \;\;\;\; z \mapsto \zeta(z) = z \otimes z^{\dagger}, 
$$
one can define a map $\iota : \C^2 \rightarrow C^+ \subset \R^{3,1}$.

Following this construction, one can associate {\em two} null vectors 
$$
\iota(\xi) = \frac{1}{2} (1,\mathbf{n}) \;\;\;\; \mbox{and} \;\;\;\; \iota(J \xi) = \frac{1}{2} (1, - \mathbf{n}), 
$$
to a given coherent state $\xi$. From these two vectors, one can construct the space-like bivector
\be
\label{northpole}
b = 2 *  \iota(J \xi) \wedge \iota(\xi),
\ee
where the star $*$ is the Hodge operator on $\Lambda^2(\R^{3,1})$.
Regarded as an anti-symmetric $4 \times 4$ matrix, $b$ is given explicitly by
\be
b = * \left[ \begin{array}{cccc} 0 & n^1 & n^2 & n^3 \\ - n^1 & 0 & 0 & 0 \\ -n^2 & 0 & 0 & 0\\ -n^3 & 0 & 0 & 0 \end{array} \right]. \nn
\ee
Thus, to every coherent state $\xi_{ab}$ of the asymptotic data, one can associate a space-like bivector $b_{ab}$. In fact, the critical point equations \eqref{cL} carry a richer geometric structure. This geometry is made transparent by acting with a Lorentz transformation on the bivectors $b_{ab}$ and defining the space-like bivectors
\be
\label{bivectorsL}
B_{ab} = k_{ab} \, \hat{X}_a \otimes \hat{X}_a \, b_{ab},
\ee
where $\hat{X}_a$ is the $\SO(3,1)$ element corresponding to $\pm X_a$ in $\SL(2,\C)$. 

These bivectors satisfy the bivector geometry conditions. 
They are {\em simple} and {\em cross-simple} by construction. The normal appearing in the cross-simplicity condition for tetrahedron $a$ is here a future pointing vector $F_a$ in the future hyperboloid $H_3^+$. This vector is the image of the reference vector $\mathcal{F} = (1,0,0,0) \in H_3^+$ under the action of $X_a$, that is, $\gamma_L(F_a) = X_a X_a^{\dagger}$.
Furthermore, the constructed bivectors satisfy {\em closure} and {\em orientation}
because of the closure condition \eqref{closure} and the spinor equations \eqref{Lo} satisfied by the critical points. To show the orientation equation, one uses the action of $J$ on $\SL(2,\C)$ given by $J X J^{-1} = (X^{\dagger})^{-1}$, for all $X$ in $\SL(2,\C)$. Under the assumption of non-degeneracy of the boundary data, these bivectors also satisfy a {\em tetrahedron} condition. Finally, the critical points determine normals $F_a$ which are either such that at least three out of the five normals are linearly independent, or such that all are pointing in the same direction. In the first case, which occurs when the solution $\{X_a\}$ does not lie in the $\SU(2)$ subgroup stabilising $\mathcal{F}$, the corresponding bivectors satisfy the {\em non-degeneracy} condition of a Minkowskian bivector geometry. 

This implies that, if the boundary data is that of a 4d Minkowskian, non-degenerate $4$-simplex, the bivectors \eqref{bivectorsL} are equal to the bivectors of (either one of) an inversion-related pair of geometric $4$-simplexes in $\R^{3,1}$, up to a sign.
Therefore, a solution to the critical point equations \eqref{veg} is equivalent to a geometric $4$-simplex in Minkowski space with spacelike tetrahedra, up to inversion. 

\subsection{Classification of the solutions}

The classification of the solutions to the critical point equations \eqref{veg} and \eqref{Lo} depends on the boundary data. In this paper, we restrict our attention to geometric boundaries of Euclidean and Lorentzian $4$-simplexes. If the boundary data is  that of a Lorentzian $4$-simplex, the critical point equations \eqref{veg} for the Ooguri and Euclidian EPRL models admit no solutions, while one can show that the equations \eqref{Lo} for the Lorentzian EPRL model admit two parity-related solutions. If the boundary data is that of an Euclidean $4$-simplex, the critical point equations \eqref{veg} admit two distinct solutions and, surprisingly,  the same is true for the Lorentzian critical point equations \eqref{Lo}. This is due to the following fact. If the group elements $X_a$ in $\SL(2,\C)$ are restricted to the unitary subgroup $\SU(2)$, the two spinor equations \eqref{Lo} collapse to a single equation which is precisely \eqref{veg} in the spinor representation. 

\subsection{Asymptotic formulae}

For geometric boundary data, we have a distinguished boundary state, the Regge state, defined by 
\eqref{Regge}. We now look at the two types of geometric boundaries considered in this paper with the boundary state given by a Regge state.

\subsubsection{Ooguri and Euclidean EPRL models}

In the spinor representation, the critical point equations \eqref{veg} involve a phase
\be
X_a^{\dagger} X_b \, \xi_{ba} = e^{i \phi_{ab}} J \xi_{ab}.
\ee
This implies that the $15j$ action evaluated at a critical point yields
\be
\label{critO}
S_{k,\bn} = 2 i \sum_{a < b} k_{ab} \phi_{ab}.
\ee

\paragraph{4d Lorentzian boundary.}
For these types of boundaries, there are no solutions to the critical point equations and the $15j$ symbol goes to zero asymptotically faster than any polynomial of $\lambda$. The same applies to the Euclidean EPRL amplitude.

\paragraph{4d Euclidean boundary.}
 In this case, there are two inequivalent  solutions to the critical point equations $\{X_a^+\}$ and $\{X_a^-\}$, the corresponding phases being noted $\phi_{ab}^{\pm}$. Using the fact that these two sets are solutions and coupling the resulting two equations leads to the following eigenvalue equation
$$
E_{ab} \, \xi_{ab} = e^{i(\phi_{ab}^+ - \phi_{ab}^-)} \, \xi_{ab}, \;\;\;\; \mbox{with} \;\;\;\; E_{ab} = X_a^{-\dagger} X_b^{-} X_b^{+\dagger} X_a^+. 
$$
Solving this equation for $E_{ab}$ and comparing it to the definition of the Euclidean dihedral angle $\Theta_{ab}^E$ for the triangle $ab$
$$
\cos \Theta_{ab}^E := N_a \cdot N_b = \frac{1}{2} \tr E_{ab},
$$
one can show that
$$
| \phi_{ab}^+ - \phi_{ab}^- | = \Theta_{ab}^E.
$$
In fact, one can solve the sign ambiguity completely; the sign between the two angles is controlled by the sign relating the bivectors \eqref{bivectorsE} to the bivectors of the geometric $4$-simplex. Fixing a choice of $\pm$ labels consistent with the relative sign, and using the canonical choice of phase for the boundary state, one arrives at the conclusion that
$$
\phi_{ab}^{\pm} = \pm \frac{1}{2} \Theta_{ab}^E,
$$
up to multiples of $\pi$ that play no role once exponentiated.

Taking this into account when evaluating the action \eqref{critO}, we can write down the asymptotic formula describing the asymptotic behavior of the Ooguri model. The asymptotic formula has two terms, corresponding to the two solutions, and is given by
\be 
15j(\lambda k,\bn) \sim \left( \frac{1}{\lambda} \right)^6 \left[ N_+ \exp \left( i \lambda \sum_{a < b} k_{ab} \Theta_{ab}^E \right) + N_- \exp \left(- i \lambda \sum_{a < b} k_{ab} \Theta_{ab}^E \right) \right],
\ee
where $N_{\pm}$ are constants that do not scale.

The asymptotic behaviour of the Euclidean EPRL model is obtained by the taking the unbalanced square of the above formula. The result reads
\be 
A^{E}_{\sigma}(\lambda k,\bn)\sim \left( \frac{1}{\lambda} \right)^{12} \left[ 2 N_{\mbox{{\tiny$+-$}}} \cos \left( \lambda \gamma \sum_{a < b} k_{ab} \Theta_{ab}^E \right) + N_{\mbox{{\tiny$++$}}} \exp \left( i \lambda \sum_{a < b} k_{ab} \Theta_{ab}^E \right)
+ N_{\mbox{{\tiny$--$}}} \exp \left(- i \lambda \sum_{a < b} k_{ab} \Theta_{ab}^E \right) \right],
\ee
where the constants $N_{\mbox{{\tiny$+-$}}}$, $N_{\mbox{{\tiny$++$}}}$ and $N_{\mbox{{\tiny$--$}}}$ do not scale.

\subsubsection{Lorentzian EPRL model}

The action at the critical point has vanishing real part and we are left with the imaginary part
\be
\label{critL}
S_{k,\bn} = i \sum_{a<b} p_{ab} \, \ln \, \frac{\parallel \z_{ba} \parallel^2}{\parallel \z_{ab} \parallel^2} + 2 n_{ab} \, \theta_{ab}.
\ee

\paragraph{4d Lorentzian boundary.}

In this case, the critical points determine a non-degenerate $4$-simplex with Lorentzian metric, up to inversion. Considering a such solution, one can couple the two spinor equations \eqref{Lo} and obtain the eigenvalue equation
$$
L_{ab} \, \xi_{ab} = e^{r_{ab}} \, \xi_{ab}, \;\;\;\; \mbox{with} \;\;\;\; L_{ab} = X_a^{-1} X_b X_b^{\dagger} (X_a^{\dagger})^{-1},
$$
and $e^{r_{ab}} =  \parallel \z_{ba} \parallel^2 / \parallel \z_{ab} \parallel^2$. Solving this equation for the Hermitian matrix $L_{ab}$ leads to the identification
$$
| r_{ab} | = |\Theta_{ab}^L|,
$$
where $\Theta_{ab}^L$ is the Lorentzian dihedral angle associated to the triangle $ab$ defined as the intersection of the two hyperplanes determined by $F_a$ and $F_b$. This triangle is a thick wedge in the terminology of \cite{Foxon} which implies that the corresponding dihedral angle is defined, up to a sign, by
$$
\cosh \Theta_{ab}^L := - F_a \cdot F_b = \frac{1}{2} \tr L_{ab}.
$$
As in the Euclidean case, the sign ambiguity can be resolved and, with the canonical choice of phase for the boundary state, one can show that the action at the critical points is given by \eqref{critL} with
$$
\ln \, \frac{\parallel \z_{ba} \parallel^2}{\parallel \z_{ab} \parallel^2} = \Theta_{ab}^L, \;\;\;\; \mbox{and} \;\;\;\; \theta_{ab} = 0, \pi.
$$
To each solution one can associate a second solution corresponding to a parity related $4$-simplex and, consequently, the asymptotic formula has two terms. It is given, up to a global sign, by the expression
\be 
A^{L}_{\sigma}(\lambda k,\bn)\sim \left( \frac{1}{\lambda} \right)^{12} \left[ N_+ \exp \left( i \lambda \gamma \sum_{a < b} k_{ab} \Theta_{ab}^L \right) + N_- \exp \left(- i \lambda \gamma \sum_{a < b} k_{ab} \Theta_{ab}^L \right) \right],
\ee
where $N_{\pm}$ are constants that do not scale. 

\paragraph{4d Euclidean boundary.}

If the boundary data is that of an Euclidean $4$-simplex, we have seen that there exist non-trivial critical points. There are two $\SU(2)$ solutions to the critical  equations for these types of boundaries which, together, build an Euclidean bivector geometry. The asymptotics are given by
\be 
A^{L}_{\sigma}(\lambda k,\bn)\sim \left( \frac{1}{\lambda} \right)^{12} \left[ N_+ \exp \left( i \lambda \sum_{a < b} k_{ab} \Theta_{ab}^E \right) + N_- \exp \left(- i \lambda \sum_{a < b} k_{ab} \Theta_{ab}^E \right) \right].
\ee

\section{Conclusion}

In this paper, we have presented results on the asymptotic behaviour of four-simplex amplitudes for the Ooguri model and the Euclidean and Lorentzian EPRL models. We used stationary phase methods applied to integral formulations of the amplitudes expressed in the coherent state basis. 
The asymptotic formulae are given, in all three cases, by simple functions of the Regge action for the $4$-simplex geometry. Note that as a corollary of our results, the asymptotics of the EPR and FK models \cite{LQG,EPR,Pereira,Livine} can be immediately derived. 


\section*{Acknowledgements} 
We would like to thank the organisers of the 2nd school and workshop on quantum gravity and quantum geometry (Corf\`u, 13-20/09/2009) where these results were presented.
WF acknowledges support from the Royal Commission for the Exhibition of 1851 and from the Emmy Noether grant ME 3425/1-1 of the German Research foundation (DFG).

\end{document}